\documentclass[11pt]{article}
\usepackage{mikemoriond,epsfig}

\begin{document}
\vspace*{4cm}
\title{NEWS FROM HERWIG}

\author{ M.H. SEYMOUR }

\address{School of Physics \& Astronomy, University of Manchester, and\\
Theoretical Physics Group (PH-TH), CERN, CH-1211 Geneva 23,
Switzerland}
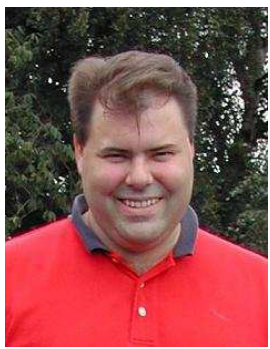
\maketitle\abstracts{I review the current status and future plans of
the HERWIG collaboration.}

\vspace*{-1.5em}
\section{Introduction}
\vspace*{-0.5em}

HERWIG\cite{Corcella:2000bw} is a Monte Carlo event generator for
simulation of hadronic final states in lepton--lepton, lepton--hadron
and hadron--hadron collisions.  It incorporates important colour
coherence effects in the final state\cite{Marchesini:1983bm} and initial
state\cite{Marchesini:1987cf} parton showers, as well as in heavy quark
processes\cite{Marchesini:1989yk} and the hard process
generation\cite{Ellis:1986bv}.  It uses the cluster\cite{Webber:1983if}
hadronization model and a cluster-based simulation of the underlying
event\cite{Marchesini:1988hj}.  While earlier
versions\cite{Marchesini:1991ch} concentrated on QCD and a few other SM
processes, recent versions contain a vast library of
MSSM\cite{Moretti:2002eu} and other BSM processes.  A review of current
Monte Carlo event generators including HERWIG can be found
in~\cite{Dobbs:2004qw}.

We are currently in a period of intense activity, finalizing the HERWIG
program and writing a completely new event generator, HERWIG++.  In this
very short contribution, I can do little more than mention the areas of
progress and provide references to sources of more details.

\vspace*{-0.5em}
\section{HERWIG version 6.5}
\vspace*{-0.5em}

HERWIG version 6.5 was released\cite{Corcella:2002jc} in October 2002.
Its main new features were an interface to the Les Houches Accord event
format\cite{Boos:2001cv}, the hooks needed by the MC@NLO
package\cite{Frixione:2002ik} and various bug fixes and minor
improvements.  It was advertised as the final fortran version of HERWIG
before work switched to HERWIG++.

Despite this, the period since then has seen intense development with
several new subversion releases and new features, most notably version
6.505, which featured an improved interface to the Jimmy generator for
multiparton interactions, which I will discuss in more detail shortly.
The most recent version is 6.507, which can be obtained from the HERWIG
web site\cite{herwig}.
\pagebreak[3]

Development of fortran HERWIG is now slowing, and the only new feature
still foreseen is the implementation of matrix element corrections to
the production of Higgs bosons, both SM and MSSM, preliminary versions
of which have been discussed in~\cite{Corcella:2004fr}.  Beyond this,
the HERWIG collaboration has made a commitment to all running (and
ceased) experiments to support their use of HERWIG throughout their
lifetimes.  Due to lack of manpower, making the same promise to the LHC
experiments would divert too much effort away from support of HERWIG++,
and we will only support their use of HERWIG until we believe that
HERWIG++ is a stable alternative for production running.

\vspace*{-0.5em}
\section{Jimmy}
\vspace*{-0.5em}

Early versions of the Jimmy model\cite{Butterworth:1996zw} generated jet
events in photoproduction using a multiparton interaction picture.  The
recent update\cite{jimmy} enables it to work efficiently as a generator
of underlying events in high $E_T$ jet events and other hard processes
in hadron--hadron collisions for the first time.  For a given pdf set,
the main adjustable parameters are \texttt{PTJIM}, the minimum
transverse momentum of partonic scattering, and \texttt{JMRAD(73)},
related to the effective proton radius.  Varying these one is able to
get a good description of the CDF data\cite{Affolder:2001xt} and other
data held in the JetWeb database\cite{Butterworth:2002ts} that are
sensitive to underlying event effects in hard process events.  However,
a poor description of minimum bias data in which there is no hard scale
is still obtained.  This is probably due to the fact that \texttt{PTJIM}
is a hard cutoff and there is no soft component below it; preliminary
attempts to rectify this are encouraging\cite{Borozan:2002fk}.  It is
interesting to note that with tunings that give equally good
descriptions of current data, Jimmy predicts twice as much underlying
event activity as PYTHIA at the LHC.

\vspace*{-0.5em}
\section{HERWIG++}
\vspace*{-0.5em}

The HERWIG program is now more than ten times the size it was when it
was designed and is maintained by a collaboration of about ten authors.
Its structure has become too unwieldy to maintain reliably and is too
rigid to incorporate many of the physics improvements that have occurred
to us recently.  We therefore took the decision to write a completely
new event generator, HERWIG++\cite{Gieseke:2003hm}, retaining the main
features of HERWIG, angular ordering and cluster hadronization, but with
a completely new design offering more flexible and scalable development,
with the aim to have a reliable product throughout the lifetime of the
LHC experiments.

\vspace*{-1em}
\paragraph{Use of ThePEG}

At around the same time, the developers of PYTHIA took a similar
decision\cite{Lonnblad:1998cq} and started work\cite{Bertini:2000uh} on
PYTHIA~7, a replacement for PYTHIA.  As part of this project, an
extremely powerful framework for the administration of event generation
was developed\cite{Lonnblad:2003wz}.  In order for this to be used by
other Monte Carlo packages, it has been separated off as an independent
Toolkit for high energy Physics Event Generation,
ThePEG\cite{Lonnblad:2004bb}.  HERWIG++ is based on ThePEG, which offers
a number of advantages: the administrative overhead is shared, while
retaining completely independent physics implementations; users only
need to learn to use one framework to use several different event
generators; and can (with care!) mix modules from different event
generators simply by selecting the appropriate components from the
toolkit.

\vspace*{-1em}
\paragraph{Hard Interactions}

A small library of basic $2\to2$ processes is built in to ThePEG, a few
more will be implemented in HERWIG++ using a new HELAS-like structure
that has already been implemented, but we do not foresee ever developing
a hard process library to the extent that we did in HERWIG.  Instead,
the plan is to provide a clean interface to external codes to generate
the majority of hard processes: interfaces to AMEGIC++ and a Les Houches
accord file reader already exist.  Of course MC@NLO- and CKKW-style
matrix element+parton shower matchings are also planned.

\vspace*{-1em}
\paragraph{Parton Showers}

A completely new parton shower algorithm has been
designed\cite{Gieseke:2003rz}, based on the quasi-collinear
limit\cite{Catani:2002hc}, which gives a smooth suppression of forward
radiation from massive partons, rather than a sharp dead-cone as in
HERWIG.  It also has advantages for matrix element matching, as it gives
a smooth coverage of the soft limit, with the emission region from
colour-connected jet pairs just touching, with no overlap or missing
region.  The new algorithm has been used for an interesting study of the
theoretical uncertainties in Sudakov form factors\cite{Gieseke:2004tc}.

\vspace*{-1em}
\paragraph{Cluster Hadronization}

The hadronization model is largely a rewrite, but with an improved
treatment of the baryon sector, inspired by, but slightly different
from, the one of Kupco\cite{Kupco:1998fx}.

\vspace*{-1em}
\paragraph{Phenomenology for \boldmath$\mathrm{e^+e^-}$ Annihilation}

The above components, together with a treatment of secondary decays that
for the moment is an exact copy of HERWIG's, are sufficient to give full
simulation of $\mathrm{e^+e^-}$ annihilation events.  The first
phenomenological study was made in~\cite{Gieseke:2003hm,Gieseke:2004af}.
Although not a complete parameter tune, a first attempt yields a similar
overall description of $\mathrm{Z^0}$ decays to HERWIG, with significant
improvements in the yields of identified baryons.  Particularly
significant is the description of B meson production~-- in HERWIG heavy
and light quark events could not be simultaneously well described,
always yielding tension in parameter tunings, but in HERWIG++ the
perturbative cutoff largely determines the B fragmentation function and
its best-fit value also gives a good description of light-quark event
shapes.

\vspace*{-1em}
\paragraph{Current Developments}

Work is under way on the extension to hadron collisions.  A simple model
of the underlying event exists and a new model based on Jimmy plus a
soft component is planned.  The initial state parton shower exists, but
needs further development and testing.  A complete description of simple
processes like Drell--Yan is anticipated for this summer.

A complete rewrite of the secondary decays is under way, with more
sophisticated treatments of almost all decay modes: a general treatment
of the spin structure and spin correlations; interference between
hadronic resonances and non-resonant diagrams; and specialized decayers
for important special cases.  The decay tables themselves are stored in
an external \texttt{xml} database and a direct link with the CEDAR
project\cite{Butterworth:2004mu} is planned.  At present 448 particles
with 2607 decay modes have been incorporated, loosely based on the
particle data group tables.  The new database allows the many massages
that are needed, for example to make branching fractions add up to
100\%, to be documented, with a star-rating system for how trustworthy
they are.

\vspace*{-1em}
\paragraph{Future Outlook}

We plan to complete the implementation and testing of initial state
showers this summer and release a preliminary version to the LHC
experiments.  Next will come 2-jet production in hadron collisions and
the first serious comparisons with Tevatron data.  Development will
continue on secondary decays and start on a multiparton interaction
model of the underlying event and CKKW-type matching to mutijet matrix
elements.  HERWIG++ will be a stable alternative to HERWIG for first LHC
analyses, allowing us a platform from which to further develop the
theoretical framework.

\vspace*{-0.5em}
\section*{Acknowledgments}
\vspace*{-0.5em}

I am grateful to all the authors of Jimmy, HERWIG, HERWIG++ and ThePEG
for their fruitful collaboration.  Stefan Gieseke's help in preparing
the transparencies for this presentation is particularly appreciated.

\end{document}